\pdfoutput=1

\documentclass[review]{elsarticle}

\usepackage{lineno,hyperref}
\usepackage{amsmath}
\usepackage{bm}
\usepackage{booktabs}
\usepackage{float}
\usepackage{multirow}
\usepackage{subcaption}

\usepackage{xcolor}

\begin{document}

\begin{frontmatter}

\title{CauchyCP: a powerful test under non-proportional hazards using Cauchy combination of change-point Cox regressions}

\author{Hong Zhang\corref{cor1}}
\ead{hong.zhang8@merck.com}\cortext[cor1]{Corresponding authors}
\author{Qing Li}
\author{Devan V. Mehrotra}
\author{Judong Shen\corref{cor1}}
\ead{judong.shen@merck.com}
\address{Biostatistics and Research Decision Sciences, Merck \& Co., Inc., Rahway, NJ, USA}

\begin{abstract}
Non-proportional hazards data are routinely encountered in randomized clinical trials. In such cases, classic Cox proportional hazards model can suffer from severe power loss, with difficulty in interpretation of the estimated hazard ratio since the treatment effect varies over time. We propose CauchyCP, an omnibus test of change-point Cox regression models, to overcome both challenges while detecting signals of non-proportional hazards patterns. Extensive simulation studies demonstrate that, compared to existing treatment comparison tests under non-proportional hazards, the proposed CauchyCP test 1) controls the type I error better at small $\alpha$ levels ($< 0.01$); 2) increases the power of detecting time-varying effects; and 3) is more computationally efficient. The superior performance of CauchyCP is further illustrated using retrospective analyses of two randomized clinical trial datasets and a pharmacogenetic biomarker study dataset. The R package \textit{CauchyCP} is publicly available on CRAN.    
\end{abstract}


\end{frontmatter}


\section{Introduction}
Time-to-event data are frequently encountered in randomized clinical trials, especially in the cardiovascular and oncology therapeutic areas. Typically, such data are analyzed using the Cox proportional hazards (PH) model for treatment effect estimation and the log-rank test for hypothesis testing. While both methods perform well under the assumption of proportional hazards, bias and loss of power can occur when the underlying hazards are no longer proportional. 

The scientific breakthrough of immuno-oncology in recent years has brought up unprecedented interest for time-to-event data with non-proportional hazards (NPH). Different types of non-proportionality have appeared in clinical trials, including and not limited to early/diminishing treatment effect, late/delayed treatment effect or crossing hazards due to their unique mechanisms. The first two scenarios are called quantitative NPH, where the hazard ratio is either $\geq 1$  or $\leq 1$  at all times, i.e., the direction of the treatment effect remains the same; the crossing hazards scenario is qualitative NPH, where the hazard ratio is $< 1$ at some time intervals, and $>1$ at some other times, i.e., the direction of the treatment effect changes during the course of trials. Crossing hazards is usually considered a special case for non-proportionality that requires further considerations in terms of results interpretation. Several real-world examples of clinical trials for PH~\citep{vermorken2007cisplatin, gandhi2018pembrolizumab}, for NPH with delayed effects~ \citep{miettinen1997cholesterol, cohen2019pembrolizumab} and for NPH with crossing hazards~\citep{mok2009gefitinib, sparano2010randomized, borghaei2015nivolumab, shitara2018pembrolizumab} can be found in the literature.

Multiple methods of both testing and estimation of treatment effect have been developed  under NPH. For treatment effect testing, the weighted log-rank \citep{harrington1982class}, weighted Kaplan-Meier \citep{pepe1989weighted, pepe1991weighted}, MaxCombo tests \citep{karrison2016versatile, lin2020alternative}, restricted mean survival time (RMST) method \citep{schaubel2011double, zhao2016restricted}, and the nonparametric K-sample omnibus non-proportional hazards (KONP) tests based on sample-space partition \citep{gorfine2020k} have been proposed. For treatment effect estimation, weighted hazard ratio \citep{chen2015quantifying}, piecewise hazard ratios \citep{liang1990cox, loader1991inference, goodman2011detecting}, RMST \citep{schaubel2011double, zhao2016restricted, horiguchi2018flexible}, and milestone survival at given time points \citep{chen2015milestone} have been studied. Each method has its own advantages and limitations. There is no general consensus on which method(s) should be applied to trial design and data analysis because many factors around the type of non-proportionality have an impact on the choice of method. In addition, under general multiple comparison scenarios or group sequential trial designs with multiple endpoints in clinical trials, small overall $\alpha$ levels (i.e., $\alpha \leq 0.01$) are frequently specified to control the family-wise error rate \citep{bretz2009graphical, maurer2013multiple}. It is not broadly known whether the existing NPH methods control the type I error for small $\alpha$ levels. 

The main challenge of considering NPH at trial design stage is that the existence and the exact nature of non-proportionality is not fully known. This results in much difficulty throughout the trial. It is difficult to determine which method(s) are robust to different possible scenarios of NPH for treatment effect testing, and hence can be pre-specified as an efficient analytical method for sample size calculation and power analysis during protocol development. It is also challenging to quantify the magnitude of treatment effect when the hazard ratio is time dependent. Difficulties exist not only in statistical sense but also in clinical interpretation and the best way to communicate varying estimates of treatment effect over time. In a nutshell, researchers are not only trying to overcome the difficulty of solving this non-proportionality problem mathematically, but also striving to find solutions to make results more clinically interpretable.

In this article, we propose CauchyCP, an omnibus test of Cox change-point models, to tackle these challenges. The CauchyCP method is demonstrated to have robust power against various NPH patterns while accurately controlling the type I error even at very stringent $\alpha$ levels such as $10^{-4}$. Besides an overall significance assessment, it also provides hazard ratio estimates before and after the most ``informative'' change point, as described later. In settings where a large number of hypotheses are tested, such as in pharmacogenetics applications, the proposed method also offers straightforward adjustment for covariates and reductions in computing time compared to existing methods like KONP and MaxCombo.

The remainder of the paper is organized as follows. Some of the common approaches are briefly introduced in the section~\nameref{sect.exist}. The details of the proposed method are described in the section~\nameref{sect.cauchy}. The simulation and real data analysis results are summarized in the sections~\nameref{sect.simu} and \nameref{sect.real}, respectively. Furthermore, some conclusions, the connection between CauchyCP and other existing methods and future potential research directions are discussed in the final section~\nameref{sect.disc}.

\section{Existing methods} \label{sect.exist}
Many statistical tests have been proposed in the literature to detect treatment effect signals under non-proportional hazards.
In this section, we briefly introduce some of the popular methods that will be compared to the proposed CauchyCP method in the simulation and real data analyses.
\begin{enumerate}
\item MaxCombo~\citep{karrison2016versatile, lin2020alternative} combines four Fleming-Harrington (FH) weighted logrank tests \cite{harrington1982class}: FH(0,0), FH(1,0), FH(1,1), and FH(0,1) using their minimum p-value. These configurations cover proportional hazards, early-difference, middle-difference, and late-difference, respectively. The final p-value is computed by assuming these four statistics follow a multivariate normal distribution with mean zero and estimated covariance matrix \citep{karrison2016versatile} under the null hypothesis.
\item KONP (K-sample omnibus non-proportional hazards tests) \citep{gorfine2020k} is a non-parametric test of non-proportional hazards based on sample-space partition. A permutation procedure is needed to calculate its final p-value. Several versions of KONP tests are proposed in literature~\citep{gorfine2020k}. Here we will use the robust version, i.e. KONP\_Cau, that combines the p-values from the KONP tests (based on Pearson chi-squared statistic and log-likelihood ratio statistic) and the logrank test, by the Cauchy combination test \citep{liu2020cauchy} for comparison.  
\item RMST (restricted mean survival time) \citep{schaubel2011double, zhao2016restricted} is the area under survival curve that measures the mean of survival time up to a fixed time cut-point. RMST can be used to detect treatment effect by taking the difference between the RMSTs of the two randomized arms being compared (e.g., treatment vs. placebo). The test statistic asymptotically follows a normal distribution. In the following simulation and real data analyses, we used the minimum of the largest observed time in each of the two arms as the cut-point.
\item WKM (weighted Kaplan-Meier) \citep{pepe1989weighted, pepe1991weighted} is a weighted difference of two Kaplan-Meier curves, where the weight at time $t$ is the harmonic mean of the probabilities of no censoring before time $t$. Similar to RMST, the test statistic of WKM asymptotically follows a normal distribution. 
\end{enumerate}
We refer the interested readers to Lin et. al. (2020)~\cite{lin2020alternative} for more comprehensive literature review of testing treatment effects under non-proportional hazard scenarios.

\section{CauchyCP} \label{sect.cauchy}
\subsection{Change-point Cox regression} 
We assume the hazard function to be in the following form:
\begin{align}
\label{equ.CP}
\lambda(t|Z,X) = \lambda_0(t)\exp(\alpha^\prime Z + \beta(t) X), 
\end{align}
where $\lambda_0$ is an arbitrary baseline hazard function, $Z$ is an $n\times p$ matrix of baseline covariates, $\alpha$ is a $p\times 1$ vector of their regression coefficients, $X$ is a variable of interest (such as a treatment indicator or a continuous biomarker variable of interest) and $\beta(t)$ is a function over $t$ that measures the time-varying effect of $X$. We would like to test the null hypothesis that $\beta(t)= 0, \forall t>0$. 

To model the non-proportional hazard functions and maintain the simplicity of the model, we further assume $\beta(t)$ is a stepwise function with $K\geq 0$ change point(s):
\begin{align}
\label{equ.step}
\beta(t) = \sum_{j=1}^{K+1} \beta_jI_{[t_{j-1},t_j)}(t),
\end{align}
with $t_0=0$ and $t_{K+1}=\infty$ and change points pre-specified at $t_1,t_2,...,t_{K}$. With this parameterization, the null hypothesis becomes   
\begin{align}
H_0: \beta_{j}=0, \quad{}j=1,...,K+1.
\end{align}

The change-point model (\ref{equ.CP}) is equivalent to a Cox regression model with time-dependent variables. To see this, we can write 
\begin{align}
\beta(t)X = \sum_{j=1}^{K+1} \beta_jX_j(t),
\end{align}
where $X_j(t) = I_{[t_{j-1},t_j)}(t)X$, $j=1,...,K+1$. Thus, the estimation of $\beta_j$, $j=1,...,K+1$ can be done winthin the Cox regression framework as described in recent publications\cite{thomas2014tutorial, zhang2018time,therneau2020using}.

\subsection{Cauchy combination of multiple Change-point Cox regressions} 
In practice, without adequate prior knowledge of the survival function (as typical in clinical trials), it is unrealistic to define a correct set of change points, $t_j$, $j=1,...,K$. If the change points are mis-specified, the power of detecting $\beta_j\neq 0$ may also suffer greatly. To address this issue, we propose CauchyCP, a Cauchy combination test \citep{liu2020cauchy} of multiple single change-point regression models, with the algorithm steps described as follows:
\begin{enumerate}
\item Pre-specify a sequence of candidate change points, $t_1,...,t_m$.
\item For each $t_i$, fit a single change-point model in equation (\ref{equ.CP}) with $\beta(t)=\beta_{i1}I_{[0,t_i)}(t) + \beta_{i2}I_{[t_i,\infty)}(t) $.
\item A p-value $P_i$ is calculated by conducting a likelihood ratio test to test the null hypothesis that $\beta_{i1}=\beta_{i2}=0$.
\item An omnibus test statistic is constructed as $cct=\sum_{i=1}^m\tan(\pi(0.5-P_i))/m$ and a final p-value is calculated as $P_{cct}=0.5-\tan^{-1}(cct)/\pi$.
\end{enumerate}
The idea behind the proposed CauchyCP method is that, although the majority of the candidate change points are likely mis-specified, at least one of the change-point models is close to the truth to capture the general pattern of the time-varying effect $\beta(t)$. Thus, by combining p-values of these change-point models, the treatment effect under non-proportional hazards can be adequately detected with properly controlled type I error. 

The p-value combination is conducted by the Cauchy combination test \citep{liu2020cauchy} which first transforms each p-value using a tangent function and then combines the transformed p-values through weighted averaging. The tangent transformation is specifically designed such that the asymptotic distribution of the combination statistic is standard Cauchy regardless of the correlation of the individual p-values. Therefore, type I error of the proposed method can be well controlled, as shown later in our simulation results, even under the scenario of correlated tests at small $\alpha$ levels. 

The proposed method is flexible in choice of the candidate change points. If the interest is in detecting early separation of the  non-proportional hazard functions, one might specify $t_i$ close $0$  or some relatively low percentiles of the event times to increase the power to identify desired signals. On the other hand, if we believe the effect is more likely to change at the later stage of the treatment or disease progression, the candidate change points should be specified further away from $0$. When conducting an analysis without specific interest/information of the hazard function, we can define $t_i$ to cover the whole range of the event times. In the rest of the paper, for fair comparison with existing methods like MaxCombo, we will use $t_1=0$,  and $t_2,t_3,t_4$ as the $25th, 50th, 75th$ percentiles of the event times, respectively.

\section{Simulation} \label{sect.simu}
We simulate right-censored data from piecewise exponential models. The hazard function of the control arm is assumed to be constant with $\lambda_C=0.1$. The hazard ratio of the treatment arm is defined by the following two configurations: 
\begin{align}
\text{configuration (1): }h_T(t) &= \sum_{i=1}^{p+1}[h_l + (i-1)\Delta_1] I_{[t_{i-1},t_i)}(t), \quad{} \label{equ.config1}\\
\text{configuration (2): }h_T(t) &= \sum_{i=1}^{p/2+1}[h_l + (i-1)\Delta_{21}] I_{[t_{i-1},t_i)}(t) + \sum_{i=p/2+2}^{p+1}[h^*+0.2 - (i-p/2-1)\Delta_{22}] I_{[t_{i-1},t_i)}(t), \label{equ.config2}
\end{align}
where $p$ is the number of change points, $\Delta_1 = (h_r-h_l)/p$, $\Delta_{21} = 2(h^*+0.2-h_l)/p$, $\Delta_{22}= 2(h^*+0.2-h_r)/p$, $h^*=\max(h_l, h_r)$, $h_l, h_r$ are the starting and ending hazards, respectively, and $t_i=8i/(p+1)$, $i=1,...,p$,  $t_0=0$ and $t_{p+1}=\infty$. In other words, the change points are equal steps from $0$ to $8$. In configuration (1), the corresponding hazard ratios of the treatment arm are also equal steps from $h_l$ to $h_r$. In configuration (2) the hazard ratios are first increasing from $h_l$ to $h^*+0.2$, and then decreasing to $h_r$. The total sample size is set as $N=100, 200,500$ and $1000$, following a 1:1 treatment-placebo allocation scheme. The censoring distribution independently follows $\exp(0.1)$. Under the null hypothesis, we fix $h_T(t)\equiv 0$. Under the alternative hypotheses, we choose $h_l,h_r\in (0.2,0.4,...,1.6)$ to explore different patterns of the hazard ratios over time, as well as different number of change points $p=1,2$ and $4$ under configuration (1) ($p=2$ and $4$ under configuration (2)). Some examples of the survival distributions of  formula (\ref{equ.config1}) and (\ref{equ.config2}) are plotted in Figure~\ref{fig.dist}. The actual number of the simulated patterns for power comparison are much larger than those plotted. 

The methods to be compared are CauchyCP with change points pre-specified at $t=0$ (the Cox PH model), $t=25th$, $50th$ and $75th$ percentiles of the event times, MaxCombo, KONP\_Cau, RMST and WKM. For type I error comparison, all methods except for KONP\_Cau were evaluated at $\alpha=0.05$ to $10^{-4}$ using $10^5$ repetitions under the null hypothesis. The type I error of KONP\_Cau was only evaluated at $\alpha=0.05, 0.025,$ and $0.01$ due to computational difficulty from its permutation test nature. The power was estimated using $2\times 10^3$ repetitions under the alternative hypotheses as the proportion of p-values less than $\alpha = 0.05$. The Cox PH model was also added as a benchmark in the power comparison. 

\subsection{Type I error comparison}
The type I error simulation results are summarized in Table \ref{tbl.TIE}, which shows that the proposed CauchyCP method controls the type I error at various $\alpha$ levels from $0.05$ to $10^{-4}$ and across different sample sizes. MaxCombo, RMST and WKM also control the type I error at $\alpha=0.05$; however, as $\alpha$ decreases, their type I error rates become inflated. The inflation is also more severe when sample size is small. For example, when $N=100$, the MaxCombo test generates type I error rates of about $1.9\alpha$ and $2.6\alpha$ when the nominal rates are $10^{-3}$ and $10^{-4}$, respectively. Similar inflation is observed for RMST at these two small $\alpha$ levels when sample size is small. The ability to control type I error at stringent $\alpha$ levels and across different sample sizes makes CaucyCP a more reliable test when $\alpha\leq 0.01$ is needed, such as in an $\alpha$ allocation procedure implemented in group sequential designs \cite{bretz2009graphical, maurer2013multiple} or a large-scale biomarker/genomics study where multiple markers need to be tested. Finally, we find that KONP\_Cau is able to control the type I error at $\alpha=0.05$. However, its permutation-based algorithm needs many more replicates to address small $\alpha$ levels and is thus much more time-consuming, which makes it not feasible for large-scale studies like genetic association analyses when thousands or even millions of tests must be performed.  

\begin{table}
\centering
\caption{Type I error simulation results from CauchyCP and other existing NPH methods. KONP\_Cau method was evaluated only at $\alpha\geq 0.01$ because of the computational difficulty from its permutation procedure. \label{tbl.TIE}}
\begin{tabular}{@{}ccccccc@{}}
\toprule
\small N                     & $\alpha$   & \small CauchCP & \small KONP\_Cau & \small MaxCombo & \small RMST    & \small WKM     \\ \midrule
\multirow{5}{*}{100}  & 5.0E-2 & 5.1E-2 & 4.6E-2   & 5.3E-2  & 4.9E-2 & 4.8E-2 \\
                      & 2.5E-2 & 2.6E-2 & 2.8E-2   & 2.7E-2  & 2.8E-2 & 2.8E-2 \\
                      & 1.0E-2 & 1.1E-2 & 9.5E-3   & 1.3E-2  & 1.5E-2 & 1.2E-2 \\
                      & 1.0E-3 & 1.1E-3 & -         & 1.9E-3  & 2.1E-3 & 1.2E-3 \\
                      & 1.0E-4 & 1.2E-4 & -         & 2.6E-4  & 3.5E-4 & 9.2E-5 \\ \midrule
\multirow{5}{*}{200}  & 5.0E-2 & 5.2E-2 & 4.6E-2   & 5.1E-2  & 4.6E-2 & 5.2E-2 \\
                      & 2.5E-2 & 2.7E-2 & 2.5E-2   & 2.7E-2  & 2.8E-2 & 2.8E-2 \\
                      & 1.0E-2 & 1.1E-2 & 9.5E-3   & 1.2E-2  & 1.3E-2 & 1.2E-2 \\
                      & 1.0E-3 & 1.2E-3 & -         & 1.8E-3  & 1.7E-3 & 1.3E-3 \\
                      & 1.0E-4 & 1.0E-4 & -         & 2.6E-4  & 2.3E-4 & 1.4E-4 \\ \midrule
\multirow{5}{*}{500}  & 5.0E-2 & 5.2E-2 & 4.7E-2   & 5.0E-2  & 4.5E-2 & 5.1E-2 \\
                      & 2.5E-2 & 2.6E-2 & 2.5E-2   & 2.6E-2  & 2.5E-2 & 2.7E-2 \\
                      & 1.0E-2 & 1.1E-2 & 1.0E-2   & 1.2E-2  & 1.1E-2 & 1.1E-2 \\
                      & 1.0E-3 & 1.0E-3 & -         & 1.5E-3  & 1.1E-3 & 1.0E-3 \\
                      & 1.0E-4 & 8.0E-5 & -         & 1.7E-4  & 7.0E-5 & 7.0E-5 \\ \midrule
\multirow{5}{*}{1000} & 5.0E-2 & 5.1E-2 & 4.6E-2   & 4.8E-2  & 4.4E-2 & 5.0E-2 \\
                      & 2.5E-2 & 2.6E-2 & 2.4E-2   & 2.6E-2  & 2.4E-2 & 2.5E-2 \\
                      & 1.0E-2 & 1.0E-2 & 9.6E-3   & 1.0E-2  & 1.0E-2 & 9.7E-3 \\
                      & 1.0E-3 & 1.0E-3 & -         & 1.4E-3  & 1.1E-3 & 9.7E-4 \\
                      & 1.0E-4 & 8.0E-5 & -         & 1.3E-4  & 1.0E-4 & 1.1E-4 \\ \bottomrule
\end{tabular}
\end{table}

\subsection{Power comparison}
\begin{figure}[]
\centering
\includegraphics[width=0.85\textwidth]{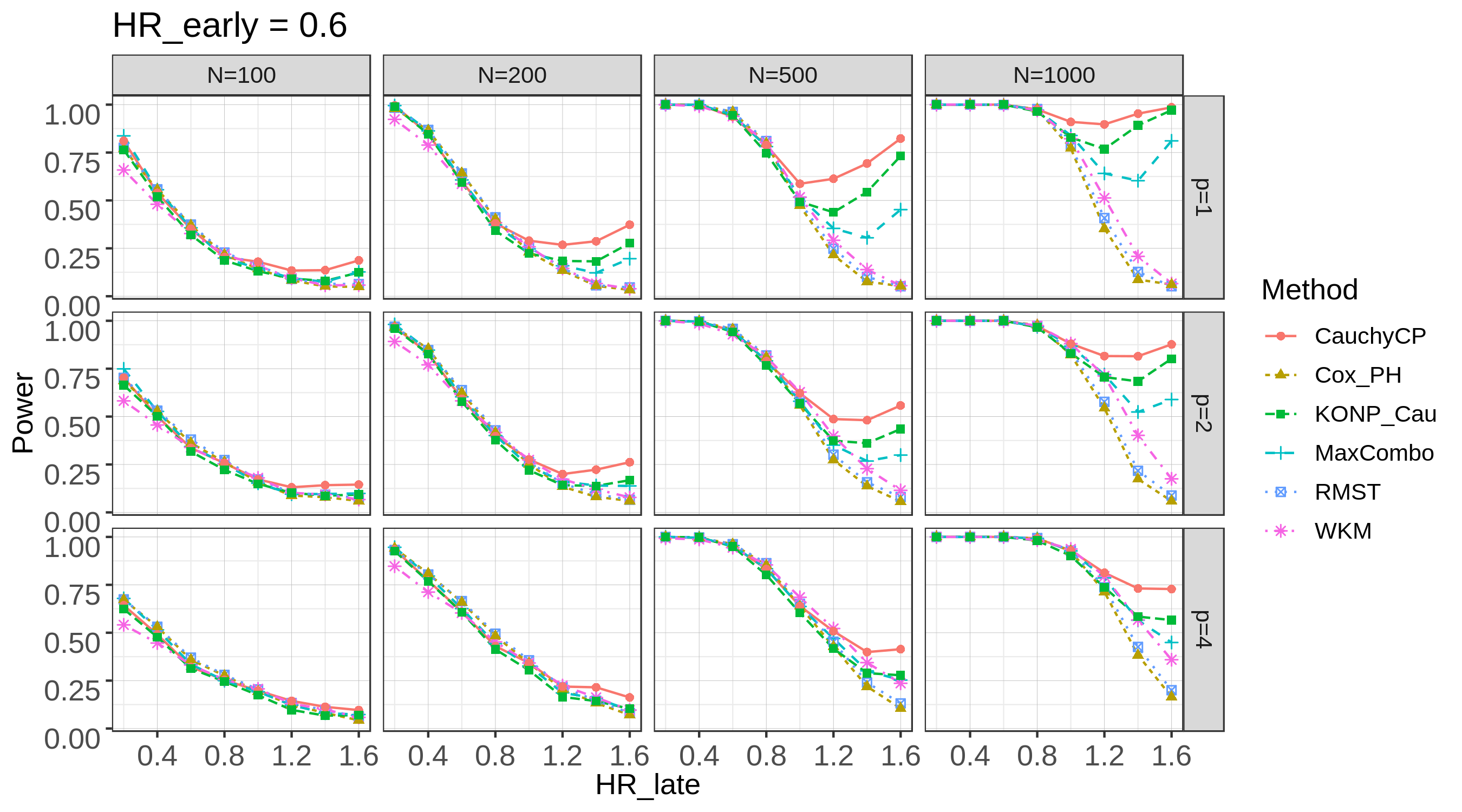}\\
\includegraphics[width=0.85\textwidth]{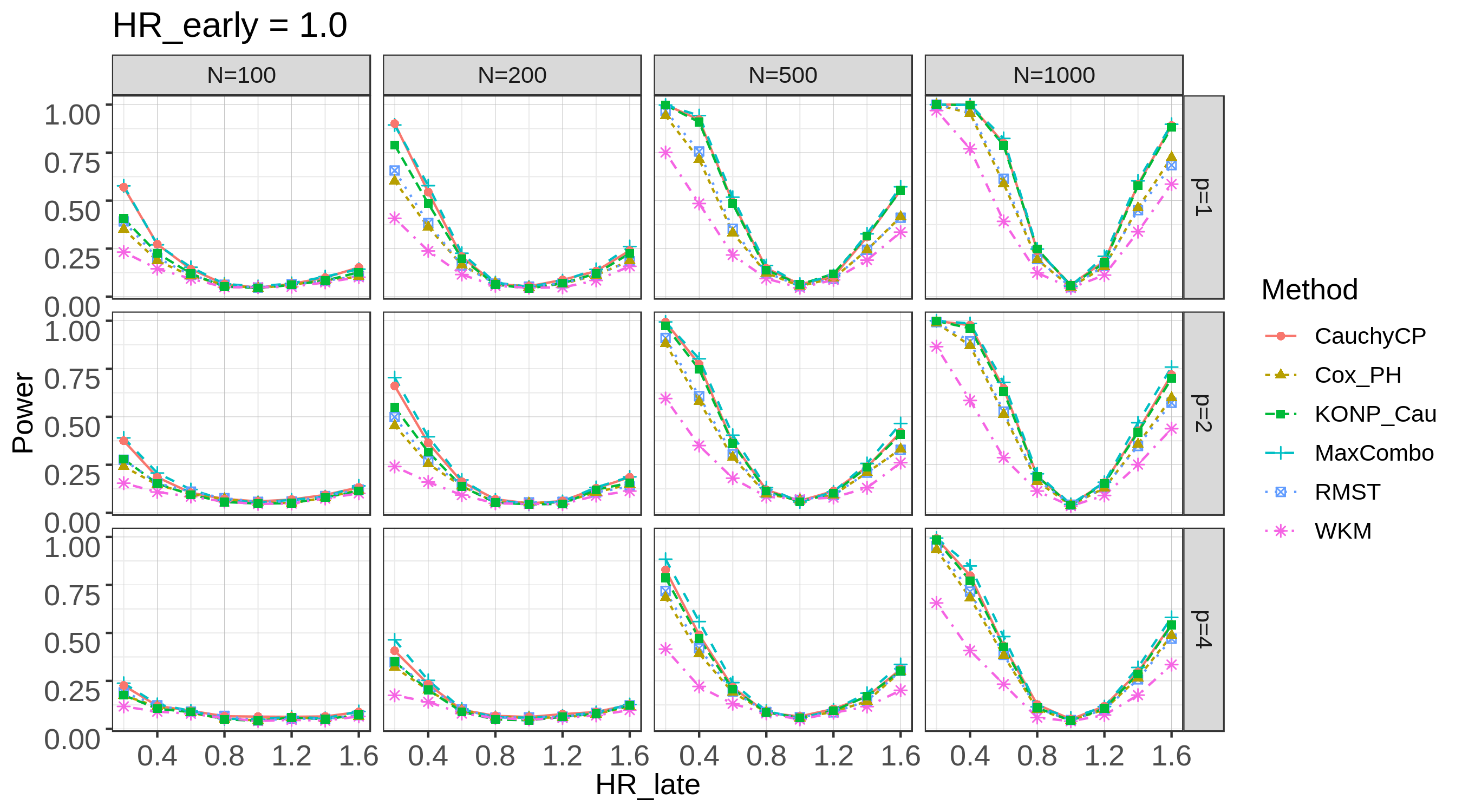}\\
\includegraphics[width=0.85\textwidth]{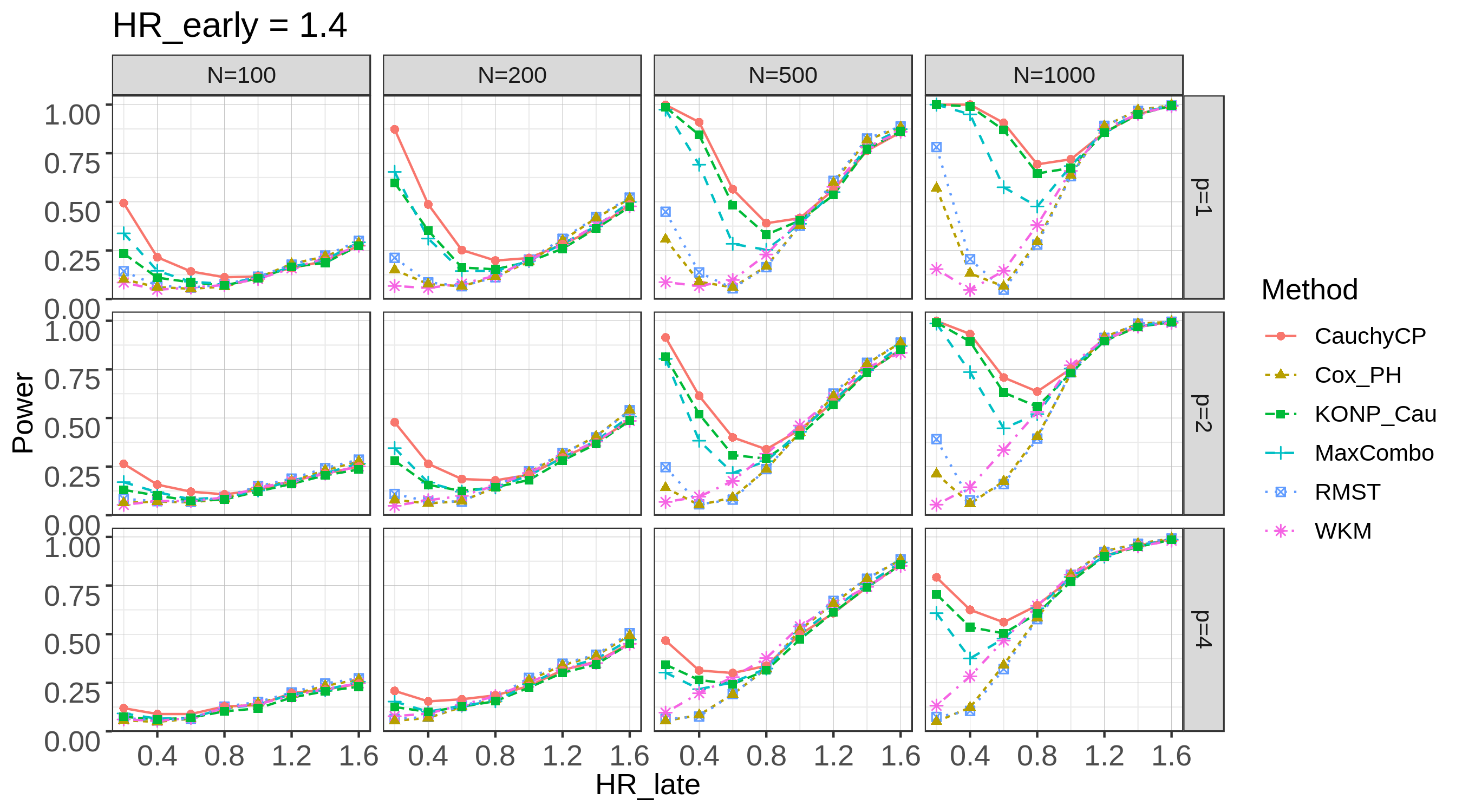}
\caption{Power comparison among the CauchyCP, Cox PH model, KONP\_Cau, MaxCombo, RMST and WKM under configuration (1). The starting hazard ratio is set as $0.6$ (upper), $1.0$ (middle) and $1.4$ (lower) and the ending hazard ratio varies from $0.2$ to $1.6$. N: sample size; p: the number of change points.  \label{fig.power}}
\end{figure}

The power simulation results under configuration (1) are summarized in Figure~\ref{fig.power}, which compares the methods by fixing the starting (earliest) hazard ratio to be $0.6$, $1$ and $1.4$, respectively. The ending hazard ratio varies from $0.2$ to $1.6$. When an early effect is present such as $h_l=0.6$ or $1.4$ and the hazard ratio changes in the opposite direction over time, the proposed CaucyCP method significantly increases the power compared to other methods. In such scenarios, the KONP\_Cau method generally performs the second best and MaxCombo comes in third place. The Cox PH model, RMST and WKM, however, are not robust against such non-proportionality as their power often drops to near zero. When there is no effect at the beginning, i.e. $h_l=1$, CaucyCP, MaxCombo and KONP\_Cau all have very similar power across different sample sizes. The power improvement of CauchyCP seems to be robust against the misspecification of change points. Even when $p=4$, we still observe a non-trivial power increase of the CaucyCP method, e.g. when $N\geq 500$ and $h_l=0.6, h_r\geq 1.2$ or $h_l=1.4, h_r\leq 0.8$. 
The power simulation results under configuration (2) are summarized in Figure~\ref{fig.power_config2}. They showed similar patterns as in configuration (1) except when $h_l=1.4$. The power advantage of the CauchyCP test is most prominent when $h_l=0.6$. CauchyCP also performs the best among the competing methods when $h_l=1$ although the margins are smaller. When $h_l=1.4$, CauchyCP's power is no longer the highest, but it is still very close to the highest. In summary, CauchyCP appears to have optimal or near optimal power compared with other existing methods regardless of the patterns of NPH. 

\subsection{Computational time comparison}
\begin{figure}[]
\centering
\includegraphics[width=0.80\textwidth]{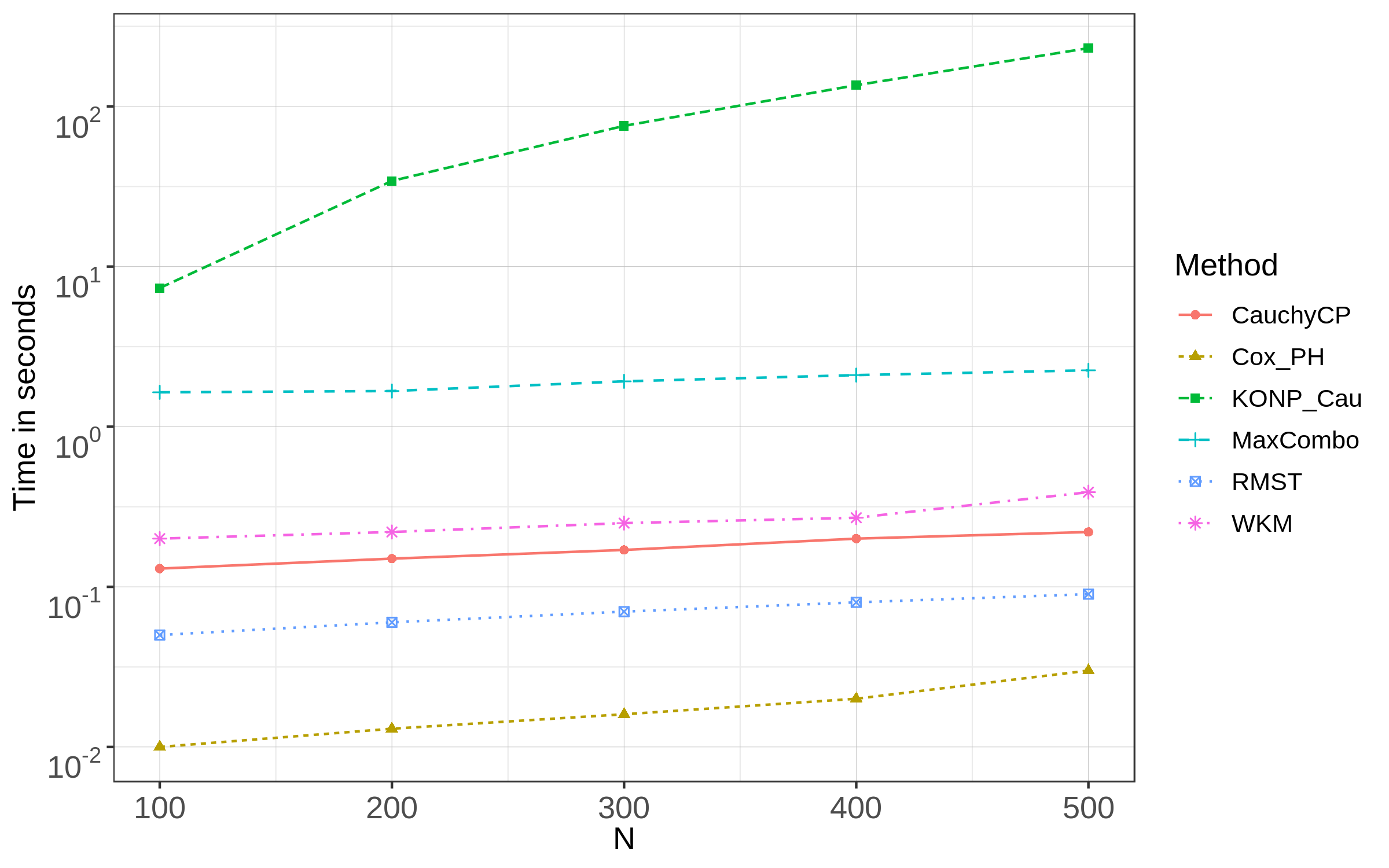}
\caption{Computational time comparison  among the CauchyCP, Cox PH model, KONP\_Cau, MaxCombo, RMST and WKM.  \label{fig.time}}
\end{figure}

We compare the computation time among the CauchyCP, Cox PH model, KONP\_Cau, MaxCombo, RMST and WKM when the sample size $N$ ranges from $100$ to $500$. The number of permutations of KONP\_Cau is set to $1000$. The computations were done on R 3.5.0, using an Intel Core-i5 8350U processor at 1.70 GHz with 16 GB of RAM. Each method was run $100$ times under the null hypothesis and the average time in seconds is plotted in Figure \ref{fig.time}. The figure shows that the computation time of KONP\_Cau is the highest and also increases dramatically as N increases, while the other methods are not that sensitive to $N$. Compared to MaxCombo, the proposed CauchyCP method reduces the computation time by almost ten folds, which makes it more suitable for large-scale analyses of time-to-event endpoints such as in biomarker studies with thousands or millions of candidate markers. 

\section{Real data analysis} \label{sect.real}
To illustrate the performance and utility of the CauchyCP method, we applied it to two clinical trial datasets followed by a pharmacogenetic biomarker study dataset. The PH assumption is violated in both the first and second datasets. More specifically, a delayed and an early treatment effects were observed in them, respectively. The first two clinical trial examples also differ in terms of the sample sizes. The third pharmacogenetics biomarker analysis example further demonstrates the utility of CauchyCP beyond clinical trials space, for example, to the general biomarker discovery space when a large number of tests exists and NPH may be present in some of them. Since we do not necessarily know the effect's direction, we conducted two-sided tests in all of our real data analyses.

\subsection*{Example 1: gastric carcinoma trial data~\cite{hess1994assessing}}
In the first example, a two-arm gastric carcinoma clinical trial was used. Ninety patients with locally advanced, non-resectable gastric carcinoma received either chemotherapy alone (N = 45) or chemotherapy plus radiation (N = 45). Eight patients are censored in each group, respectively. The Kaplan-Maier curves in Figure~\ref{fig.hess} show that there is an obvious early effect of the treatment, but such treatment effect starts diminishing at $t\approx 300$ to $t\approx 700$, after which the two curves converge. The p-value from the popular Grambsch and Therneau (GT) test~\cite{grambsch1994proportional} is $0.0034$, which further confirms the strong evidence against proportional hazards.


CauchyCP, MaxCombo and KONP\_Cau were applied to test whether there is a statistically significant treatment effect. Table~\ref{tbl.hess} shows that the classic logrank test or Cox proportional hazard regression missed this potential signal with a p-value of $0.25$. The CauchyCP test, however, is able to detect such signal with a p-value of $0.0141$. The smallest individual p-value indicates that the change point is most likely at the median event time $t=355$. This finding is consistent with the Kaplan-Maier curves which show a sudden drop in the control arm around $t=375$. The estimated hazard ratio before the change point is $2.78$ and the one after the change point is $0.61$. However, the MaxCombo test fails to identify this signal after accounting for the four FH tests being conducted (p-value $=0.0613$). KONP\_Cau returns a statistically significant result (p-value $=0.0230$) but the significance is lower than that using CauchyCP. 

The CauchyCP results indicate a qualitative interaction of two hazard functions. For such crossing hazards, the p-values from traditional methods, e.g. the logrank test, are difficult to interpret and often lead to inconsistent benefit-risk assessment of the experimental therapy. However, the ability of modeling the change points of hazard functions, as provided by the proposed CauchyCP method, offers clearer interpretation in such scenarios. 


\begin{table}[H]
\caption{Results of the gastric carcinoma data analysis using CauchyCP and other competing methods. \label{tbl.hess}}
\begin{subtable}[c]{.75\textwidth}
\caption{P-values of CauchyCP and competing methods. The number of permutations for KONP\_Cau is 1,000. \label{tbl.hess_pval}}
\begin{tabular}{@{}lllll@{}}
\toprule
CauchyCP & KONP\_Cau & MaxCombo & RMST   & WKM    \\ \midrule
0.0141   & 0.0230     & 0.0613   & 0.3119 & 0.1150 \\ \bottomrule
\end{tabular}
\end{subtable}\\
\bigskip

\begin{subtable}[c]{\textwidth}
\captionsetup{singlelinecheck = false, justification=justified}
\caption{Estimates and p-values of each change-point model in CauchyCP. \label{tbl.hess}}
\begin{tabular}{@{}lllll@{}}
\toprule
\begin{tabular}[c]{@{}l@{}}Change\\ point\end{tabular} & \begin{tabular}[c]{@{}l@{}}PH\\\quad{}\end{tabular}    & \begin{tabular}[c]{@{}l@{}}182\\ (25\%)\end{tabular} & \begin{tabular}[c]{@{}l@{}}355\\ (50\%)\end{tabular} & \begin{tabular}[c]{@{}l@{}}540\\ (75\%)\end{tabular} \\ \midrule
HR\_early                                              & 1.30   & 3.17                                                 & 2.78                                                 & 1.61 \\
HR\_late                                               & 1.30     & 0.98 & 0.61                                                 & 0.70                                                 \\
p-values                                               & 0.2570 & 0.0603 & 0.0039                                               & 0.1609 \\ \bottomrule
\end{tabular}
\end{subtable}
\end{table}

\subsection*{Example 2: hematological malignancy clinical trial~\cite{lipkovich2017tutorial}}
{In the second example, we consider a real phase III randomized oncology (hematological malignancy) clinical trial described in the literature~\cite{lipkovich2017tutorial}. $599$ patients with a hematological malignancy were randomly assigned to either an experimental therapy plus best supporting care (treatment arm, $N=303$) or best supporting care (control arm, $N=296$). The primary endpoint in the trial was overall survival.} The Kaplan-Maier curves in Figure~\ref{fig.lipk} show that there is a delayed treatment effect until $t\approx 50$, then the treatment effect seems to increase before starting to diminish from $t\approx 100$. The GT test confirms the strong evidence against proportional hazards with a p-value of $0.0023$.

The original publication~\cite{lipkovich2017tutorial} used this as an example to demonstrate that a subgroup effect may cause non-proportional hazards phenomenon. Our results show that by using CauchyCP (or other robust tests like MaxCombo), we can directly detect the treatment effect under non-proportional hazards. Table~\ref{tbl.lipk} shows that the two sided logrank test's p-value $=0.071$.
On the other hand, the CauchyCP test gives a p-value $=8.5\times 10^{-4}$ which greatly improves the statistical significance of this potential signal. The smallest p-value comes from the $25\%$ time change point at $t=83$, after which the treatment effect starts to decrease as shown in Figure~\ref{fig.lipk}. The estimated hazard ratios before and after the change point are $0.65$ and $1.12$, respectively. In this data, the MaxCombo is also able to detect the treatment effect but the p-value $(0.0089)$ is less significant compared to that from CauchyCP. The p-value (6.0E-4) from KONP\_Cau is slightly smaller than the CauchCP's, however, its computation time is significantly higher (about half an hour vs. less than 0.1 seconds).   

Similar to Example 1, the CauchyCP results demonstrate its unique capability to infer the time-varying effect. In this example, the estimated hazard ratio 0.85 from the Cox PH model could be somewhat misleading because it was assumed constant over time. CauchyCP is a better approach in this scenario as it provides a more comprehensive assessment of the relative hazard functions than the competing methods. 

\begin{table}[H]
\caption{Results of the hematological malignancy data analysis using CauchyCP and other competing methods. \label{tbl.lipk}}
\begin{subtable}[c]{0.75\textwidth}
\caption{P-values of CauchyCP and the competing methods. The number of permutations for KONP\_Cau is 10,000. \label{tbl.lipk_pval}}
\begin{tabular}{@{}lllll@{}}
\toprule
CauchyCP & KONP\_Cau & MaxCombo & RMST   & WKM    \\ \midrule
8.5E-4   &  6.0E-4    & 0.0089   & 0.1617  &  0.0111 \\ \bottomrule
\end{tabular}
\end{subtable}\\
\bigskip

\begin{subtable}[c]{\textwidth}
\captionsetup{singlelinecheck = false, justification=justified}
\caption{Estimates and p-values of each change-point model in CauchyCP. \label{tbl.lipk}}
\begin{tabular}{@{}lllll@{}}
\toprule
\begin{tabular}[c]{@{}l@{}}Change\\ point\end{tabular} & \begin{tabular}[c]{@{}l@{}}PH\\\quad{}\end{tabular}     & \begin{tabular}[c]{@{}l@{}}83\\ (25\%)\end{tabular} & \begin{tabular}[c]{@{}l@{}}177\\ (50\%)\end{tabular} & \begin{tabular}[c]{@{}l@{}}333\\ (75\%)\end{tabular} \\ \midrule
HR\_early                                              & 0.85   & 0.48 & 0.65 & 0.76                                                 \\
HR\_late                                               & 0.85    & 1.03 & 1.12 & 1.21                                                 \\
p-values                                               & 0.0711 & 2.4E-4 & 0.0024 & 0.0163 \\ \bottomrule
\end{tabular}
\end{subtable}
\end{table}

\subsection*{Example 3: pharmacogenetic biomarker study of a cardiovascular disease trial}
{To further illustrate the unique ability of the proposed CauchyCP method to control type I error when multiple tests are present, we applied the CauchyCP and MaxCombo methods to a hypothetical pharmacogenomics (PGx) genome-wide association study (GWAS) based off a real cardiovascular disease clinical trial \citep{cannon2015ezetimibe}. The objective of the PGx GWAS study was to discover genetic variants that influence efficacy of a novel treatment while using a time-to-cardiovascular-events (CV) endpoint.} 

For demonstration purposes, we randomly selected a subset of 500 subjects of European ancestry (control: 255, treatment: 245) from the original PGx GWAS data, and used all of the 95,613 genetic variants (single nucleotide polymorphisms or SNPs) on chromosome 22 with minor allele frequency $>5\%$ for association analysis. In addition to CauchyCP, only MaxCombo was applied to this analysis. KONP\_Cau is not feasible for such type of large-scale analysis due to its high computational cost. RMST and WKM are not suitable for this exploratory study because their power is sensitive to different NPH patterns and it is expected that different NPH patterns may be present marginally across the large number of genetic markers. Also, to the best of the authors' knowledge, the current implementation of WKM does not support adjustment for covariates. For RMST, the covariates adjustment is available in the R package \emph{survRM2}. However, its implementation is time-consuming, and thus not computationally feasible for GWAS analysis. Five genetic principal components were used as covariates in the CauchyCP method. They were not used in the MaxCombo method since it cannot adjust for continuous covariates. 

\begin{figure}[]
\centering
     \begin{subfigure}[b]{0.49\textwidth}
         \centering
         \includegraphics[width=1\textwidth]{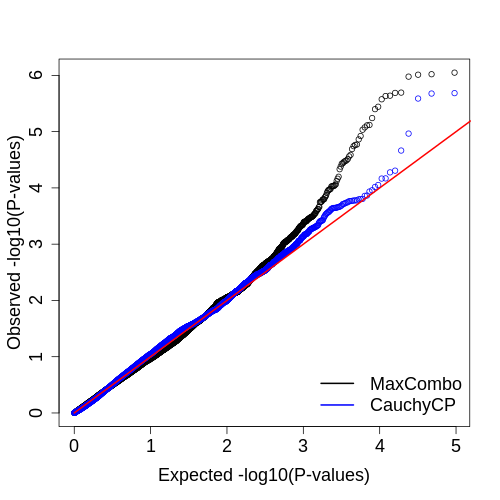}
		 \captionsetup{justification=centering}
		 \caption{}
         \label{fig.gwas_qq}
     \end{subfigure}
\hfill
     \begin{subfigure}[b]{0.49\textwidth}
         \centering
         \includegraphics[width=1\textwidth]{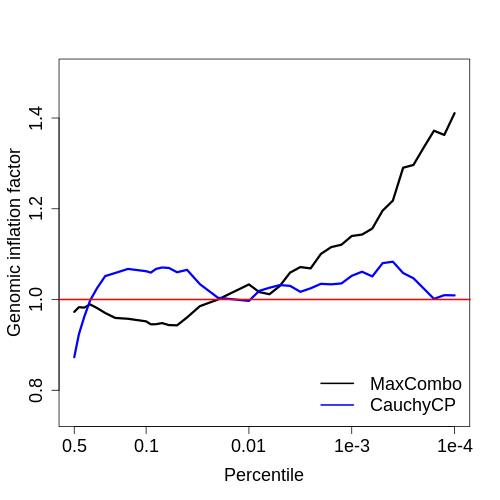}
		 \captionsetup{justification=centering}
		 \caption{}
         \label{fig.gwas_inflation}
     \end{subfigure}
         \caption{The quantile-quantile plot (left) and the genomic inflation factor plot (right) from the analysis of SNPs on chromosome 22 in a PGx GWAS study of cardiovascular events for a subset of samples in the IMPROVE-IT clinical trial.}\label{fig.gwas}
\end{figure}

We used the genomic inflation factor~\citep{devlin1999genomic} $\lambda_p$ to measure the possible inflation of type I errors at a given percentile $p$, $\lambda_p = F_{\chi_1^2}^{-1}(1-\text{pval}_p)/F_{\chi_1^2}^{-1}(1-p)$, 
where $\text{pval}_p$ denotes the $p$th percentile of the calculated p-values. A $\lambda_p$ close to $1$ indicates a well-controlled type I error rate. Figure~\ref{fig.gwas} shows that the the type I error rates of MaxCombo start to be inflated when the p-value $<0.01$. There could be two reasons for the inflation: 1) MaxCombo cannot adjust for genetic principal components thus the population stratification might inflate its type I error; 2) the p-value computation of MaxCombo is not accurate enough for stringent $\alpha$ levels as also evidenced by the type I error simulation results. On the other hand, the genomic inflation factors of the proposed CauchyCP method are around $1$, indicating a better controlled type I error rate even at small $\alpha$ levels. 

\section{Discussion} \label{sect.disc}

In the context of randomized clinical trials, 
survival analysis of time-to-event endpoints is often performed to estimate and test for the effect of an experimental drug compared to the control.
In current practice, a proportional hazards model is often assumed, and its validity is critical for the power performance of the classic log-rank test or Cox regression model. However, non-proportional hazards (NPH) are not uncommon in clinical trials, especially in recently emerging immuno-oncology trials. Some new methods are proposed to improve the power in these scenarios based on different assumptions on the NPH mechanisms, such as heterogeneous effect among subjects. 
One such example is the recently proposed 5-STAR method~\citep{mehrotra2020survival} which is an aggregation test of model-averaged accelerated failure time models applied to algorithmically defined homogeneous patient subgroups.
CauchyCP, on the other hand, assumes all subjects experience the same time-varying effect. In this sense, CauchyCP is complementary to 5-STAR based on different assumptions about the true nature of NPH. Direct comparison between these two methods is difficult because it is essentially the underlying data generating process or ground truth of the data for analysis that determines the results.

In this article, we are not trying to pinpoint the real cause of NPH. The proposed CauchyCP is flexible and robust enough to pick up the effects when the PH assumption is violated as long as the survival curves are separated enough. Our extensive simulations have shown that CauchyCP can accurately and efficiently compute the p-value even as small as $10^{-4}$, which makes it suitable for analysis of clinical trial data with small $\alpha$ levels (e.g., due to multiplicity control) and other large-scale biomarker analyses (e.g., candidate gene studies and GWAS) with large number of tests. In a general situation of testing multiple hypotheses repeatedly in time or when applying group sequential trial designs to multiple endpoints in clinical trial design and analysis, the small $\alpha$ levels are frequently specified to control the family-wise error rate \citep{bretz2009graphical, maurer2013multiple}. If NPH is anticipated in such cases, CauchyCP is a promising alternative to other existing methods in terms of better control of type I error. In addition, CauchyCP has robust power against various NPH patterns such as early/diminishing treatment effect, late/delayed treatment effect, and crossing hazards as evidenced by our simulation studies and real data analyses. Furthermore, compared with existing methods like KONP and MaxCombo, CauchyCP also enjoys the benefit of efficient adjustment of both categorical and continuous covariates.   

Another benefit of using CauchyCP is to provide extra information about the NPH pattern, which may improve the interpretation of the results. As shown in the CauchyCP algorithm, in addition to one final p-value, it also generates four change points, each with one p-value and two hazard ratio estimates (one before and one after the change point). Although we are using the Cauchy combination test, which is an aggregation of individual transformed p-values, the smallest p-value often dominates the combination statistic. Thus, it is reasonable to assume the change-point model with smallest p-value is the most informative one. As demonstrated in the first two real data analysis examples, the most informative change point is indeed visually intuitive, and the corresponding hazard ratio estimates further help the interpretation of the results.   

As one of the future research directions, we might consider including models with multiple change-points into CauchyCP. Using the Cauchy combination method and our current framework, the p-values from these more complicated models can be combined without extra difficulties in controlling the type I error.
Although our simulations have shown that the current single-change-point model is flexible enough to handle most NPH patterns, adding more complex change-point models may further improve the power, especially when prior information is available for constructing such models. 
 
CauchyCP can be also extended to conducting meta-analysis of multiple data sources in NPH scenarios. The way that we define change points allows us to meta-analyze each dataset in terms of the percentiles rather than the values of the survival time, which is helpful since the survival time of different clinical trials may be significantly different. Thus, for each pre-specified percentile, we can derive one meta p-value and two meta hazard ratios. A Cauchy combination test can be used again to aggregate the meta p-values so that the statistical significance of the meta-analysis can be assessed. We envision that such approach will have wide applications in survival analysis of multiple clinical trials and large-scale biomarker studies such as GWAS.

\section*{Acknowledgements}
The authors would like to thank Amarjot Kaur, Rachel Marceau West, Lingkang Huang, Yiwei Zhang and Aparna Chhibber for data preparation and helpful discussion.

\section*{Declaration of conflicting interests}
H.Z., Q.L., D.V.M. and J.S. are employees of Merck Sharp \& Dohme Corp., a subsidiary of Merck \& Co., Inc., Kenilworth, NJ, USA and shareholders in Merck \& Co., Inc., Kenilworth, NJ, USA.

\bibliography{mybib}

\setcounter{figure}{0}
\renewcommand\thefigure{S\arabic{figure}} 
\newpage
\section*{Supplementary Materials}

\begin{figure}[ht]
\centering
\begin{subfigure}[b]{0.825\textwidth}
\captionsetup{justification=centering}
\caption{Configuration (1)}
\centering
\includegraphics[width=0.825\textwidth]{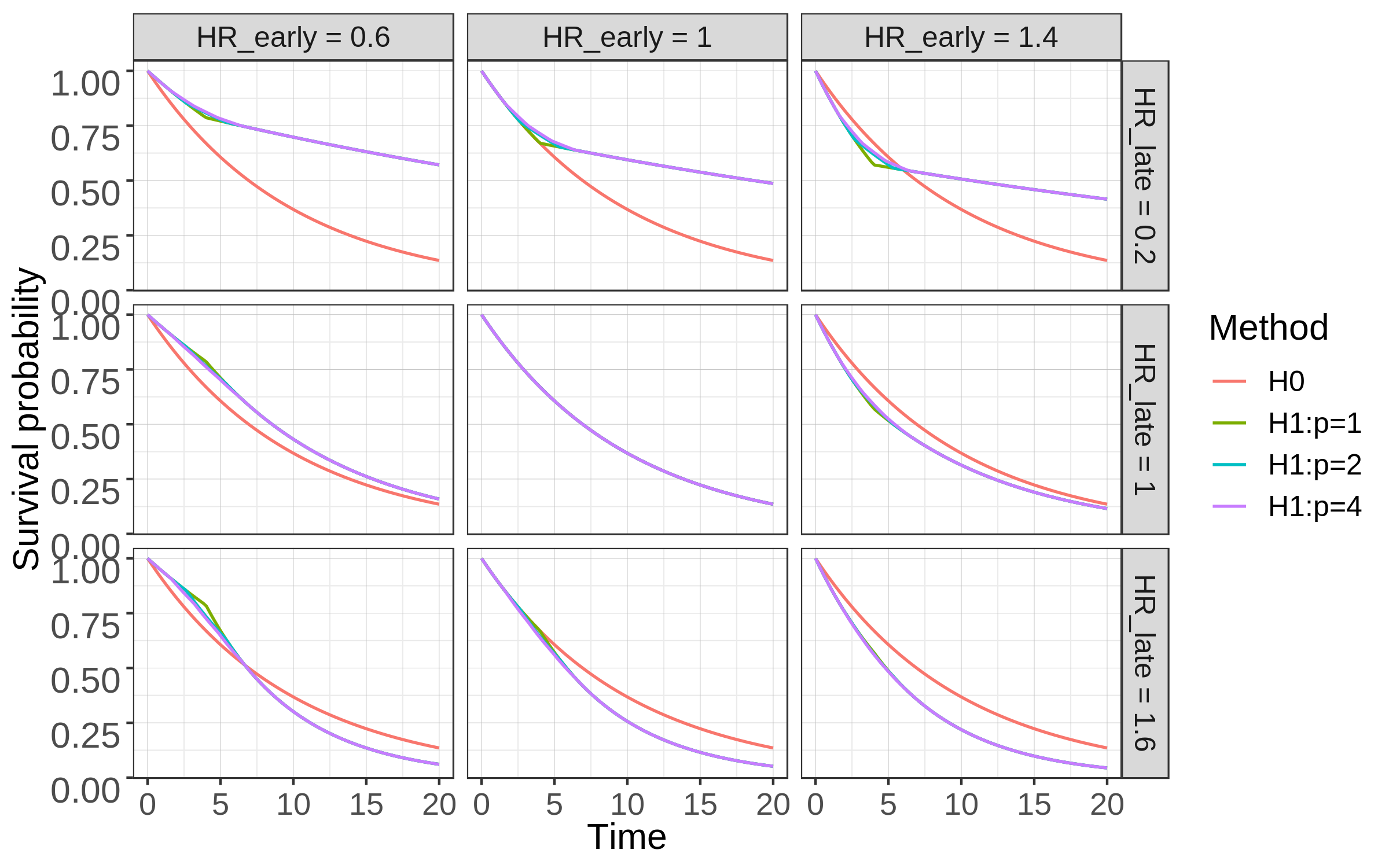}
\end{subfigure}
\bigskip

\begin{subfigure}[b]{0.825\textwidth}
\captionsetup{justification=centering}
\caption{Configuration (2)}
\centering
\includegraphics[width=0.825\textwidth]{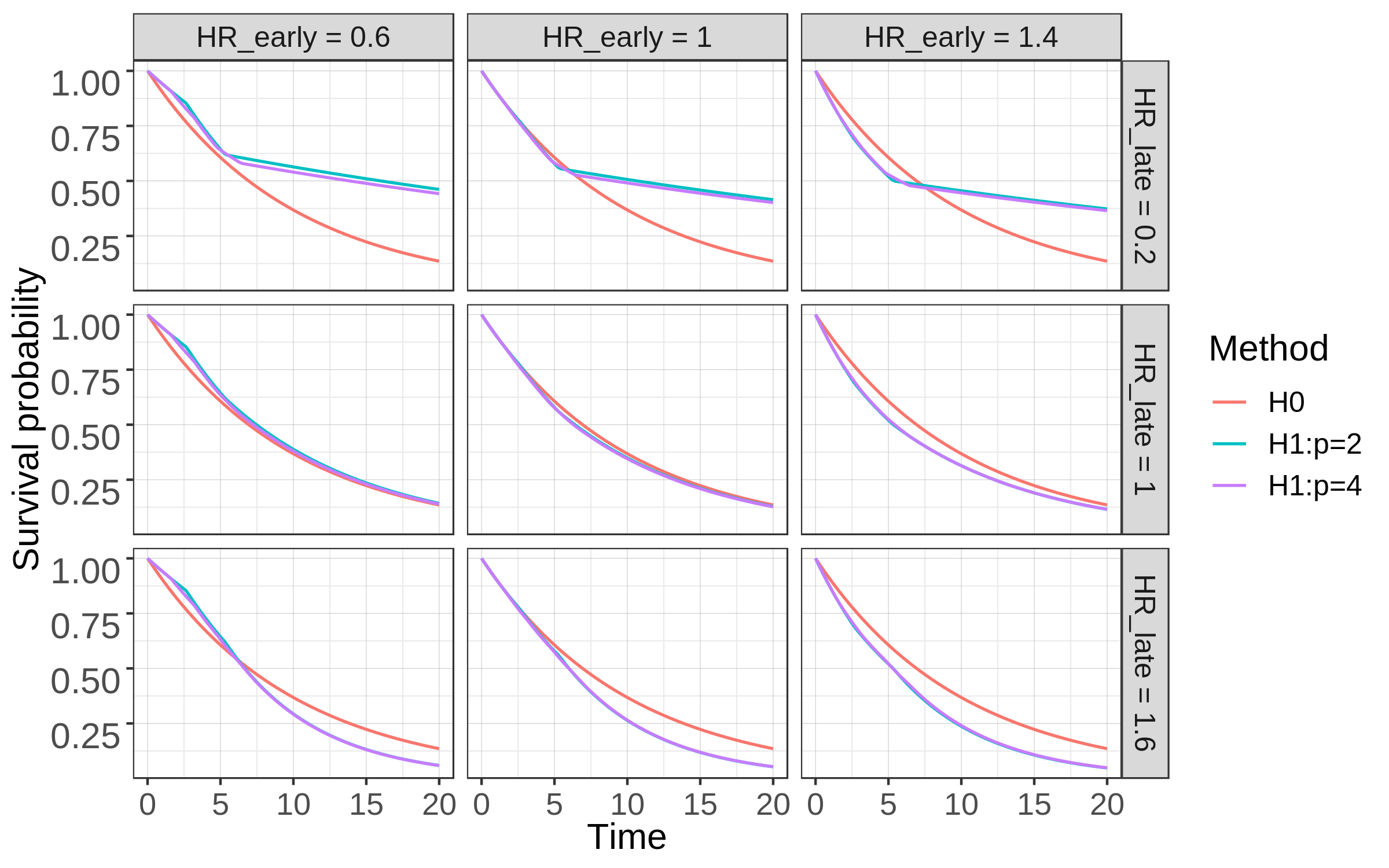}
\end{subfigure}
\caption{Survival probability of the control arm (H0) vs treatment arm (H1) under example simulation scenarios. $p$ is the number of change points. Configuration (1): formula (\ref{equ.config1}), hazard ratio monotonely changes from HR\_early to HR\_late. Configuration (2): formula (\ref{equ.config2}), hazard ratio first increases from HR\_early to $\max$\{HR\_early, HR\_late\} + 0.2 then decreases to HR\_late. \label{fig.dist}}
\end{figure}

\begin{figure}[ht]
\centering
\includegraphics[width=0.95\textwidth]{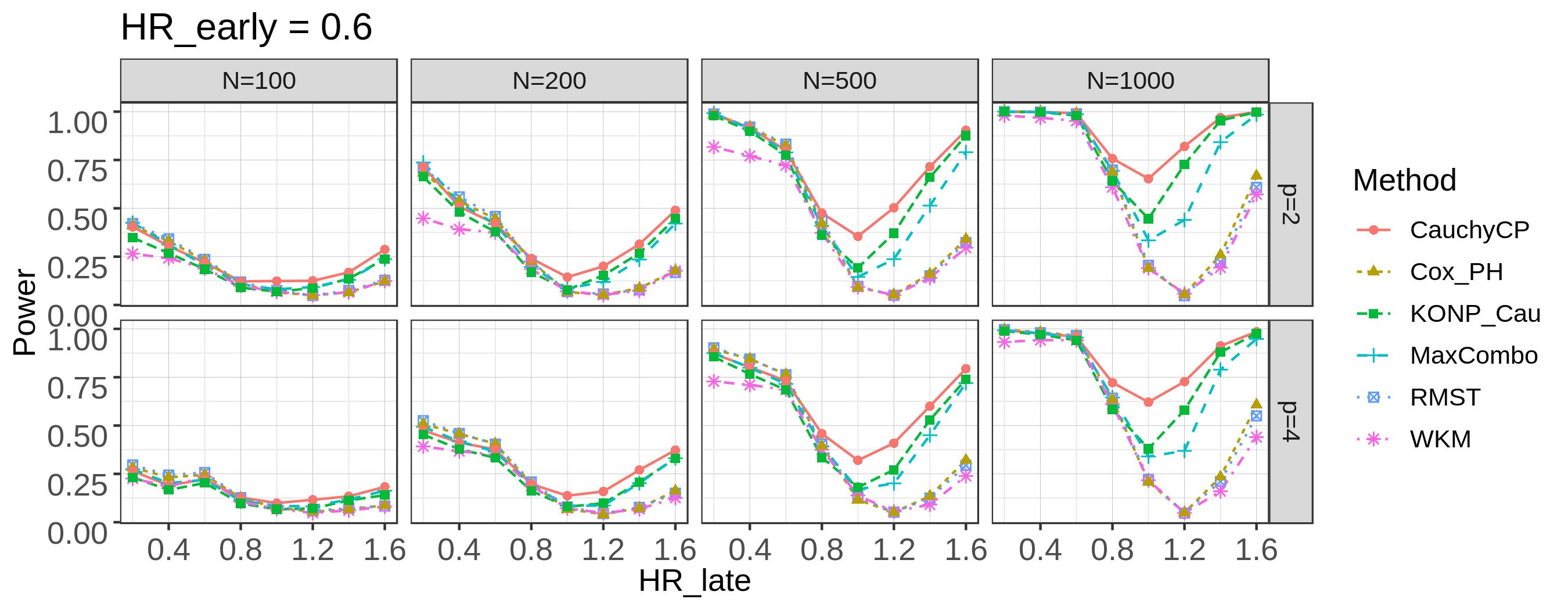}\\
\includegraphics[width=0.95\textwidth]{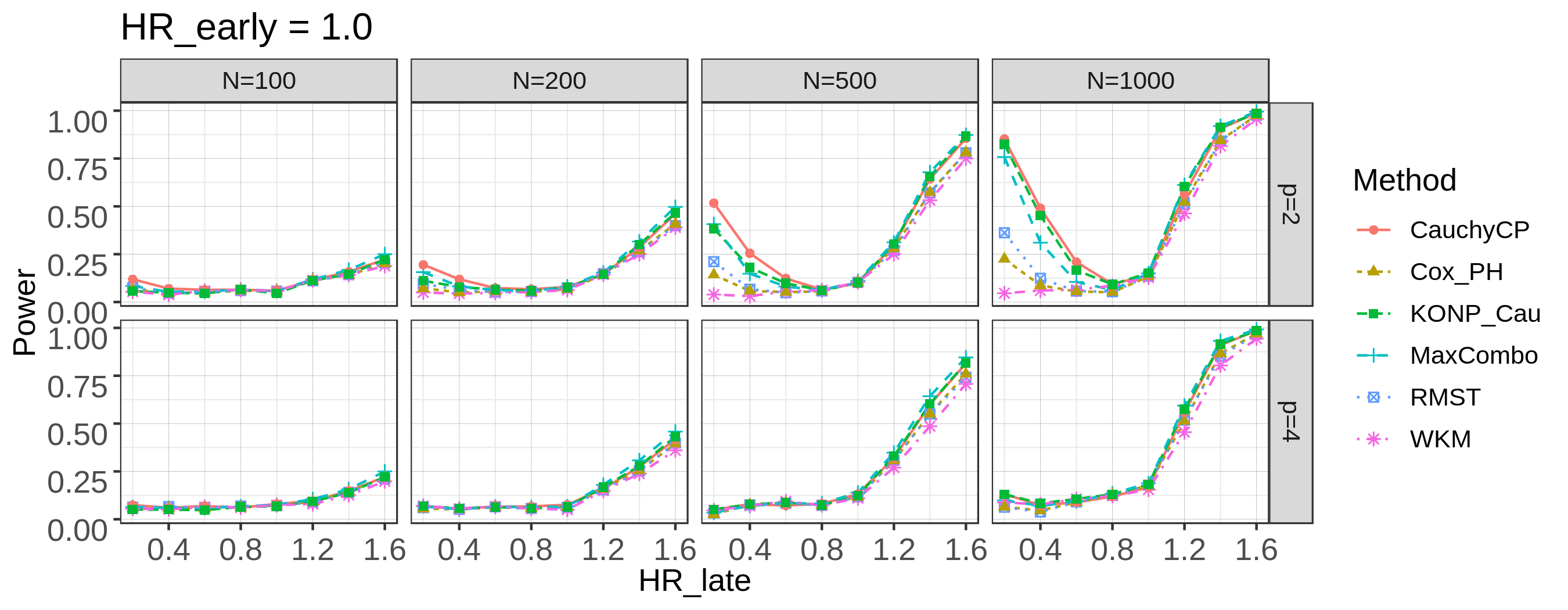}\\
\includegraphics[width=0.95\textwidth]{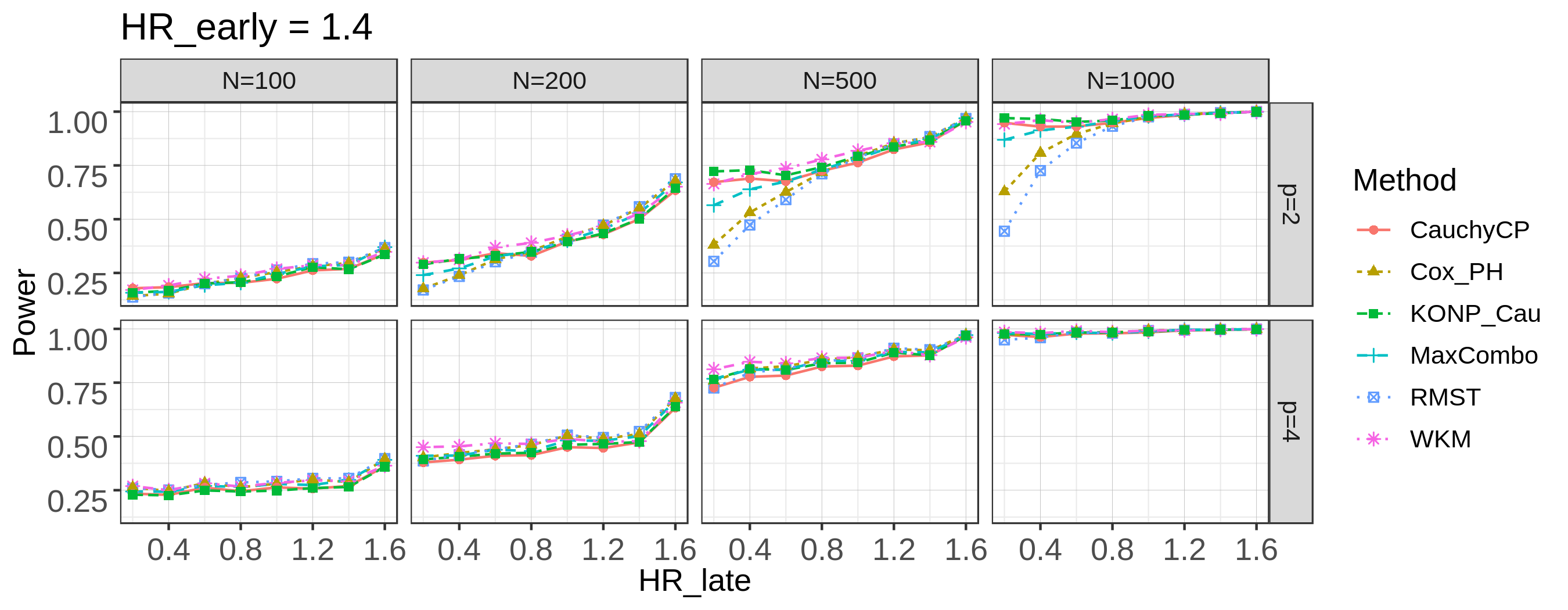}
\caption{Power comparison among the CauchyCP, Cox PH model, KONP\_Cau, MaxCombo, RMST and WKM under configuration (2). The starting hazard ratio is set as $0.6$ (upper), $1.0$ (middle) and $1.4$ (lower) and the ending hazard ratio varies from $0.2$ to $1.6$. N: sample size; p: the number of change points.  \label{fig.power_config2}}
\end{figure}

\begin{figure}[ht]
\centering
         \includegraphics[width=0.85\textwidth]{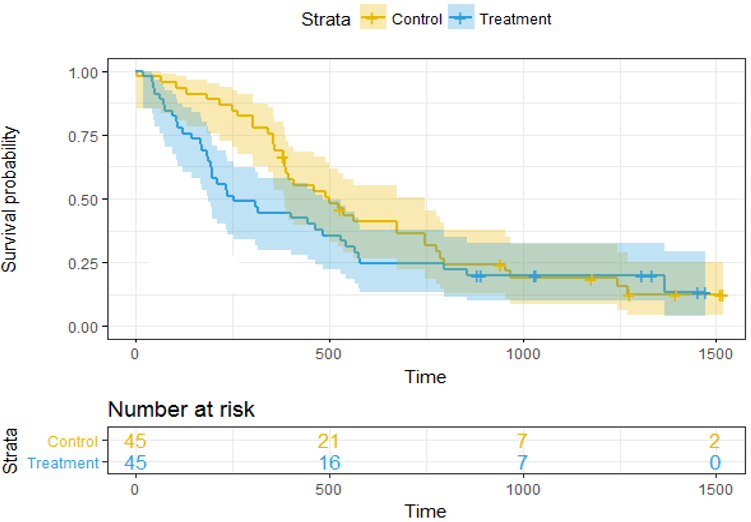}
         \caption{The Kaplan-Maier curves of the gastric carcinoma data~\cite{hess1994assessing}. Control: chemo only; Treatment: chemo + radiation. }         \label{fig.hess}
\end{figure}

\begin{figure}[ht]
         \centering
         \includegraphics[width=0.85\textwidth]{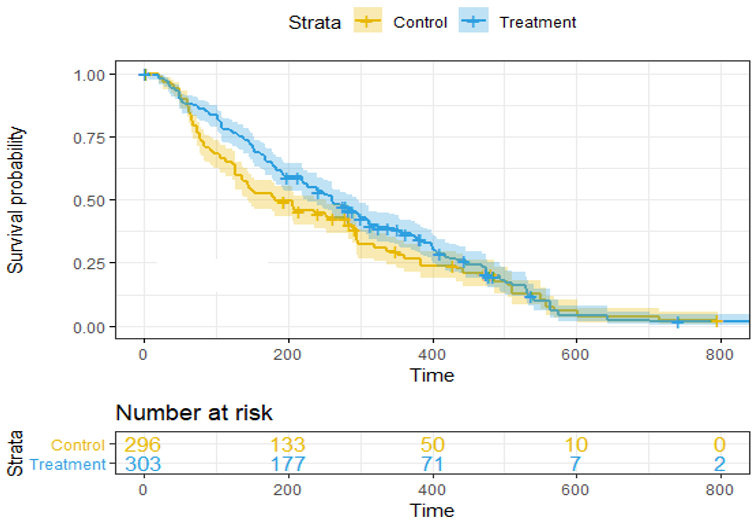}
         \caption{The Kaplan-Maier curves of the hematological malignancy data~\cite{lipkovich2017tutorial}. Control: best supporting care; Treatment: experimental therapy + best supporting care.}         \label{fig.lipk}
\end{figure}

\end{document}